\documentclass{aa}

\usepackage{graphicx}
\usepackage{txfonts}
\usepackage{color}

\usepackage[colorlinks=true, citecolor=blue, urlcolor=blue, linkcolor=blue]{hyperref}

\usepackage{natbib,twoopt}

\bibpunct{(}{)}{;}{a}{}{,} 
\newcommandtwoopt{\citeads}[3][][]{\href{http://adsabs.harvard.edu/abs/#3}%
{\citealp[#1][#2]{#3}}}
\newcommandtwoopt{\citepads}[3][][]{\href{http://adsabs.harvard.edu/abs/#3}%
{\citep[#1][#2]{#3}}}
\newcommandtwoopt{\citetads}[3][][]{\href{http://adsabs.harvard.edu/abs/#3}%
{\citet[#1][#2]{#3}}} 
\newcommandtwoopt{\citeyearads}[3][][]%
{\href{http://adsabs.harvard.edu/abs/#3}{\citeyear[#1][#2]{#3}}} 

\begin{document}

   \title{Spatial distribution of FIR rotationally excited CH$^+$ and OH emission lines in the Orion Bar PDR\thanks{\textit{Herschel} is an ESA space observatory with science instruments provided by European-led Principal Investigator consortia and with important participation from NASA.}}

   \author{A. Parikka \inst{1,2}
          \and
          E. Habart \inst{1}
          \and
          J. Bernard-Salas \inst{3}
          \and
          J. R. Goicoechea \inst{4}
          \and
          A. Abergel \inst{1}
          \and
          P. Pilleri \inst{5,6}
          \and
          E. Dartois \inst{1}
          \and
          C. Joblin \inst{5,6}
          \and
          \mbox{M. Gerin} \inst{7,8}
          \and
          B. Godard \inst{7,8}
                    }

    \institute{Institut d'Astrophysique Spatiale, Universit\'e Paris-Saclay, 91405 Orsay Cedex, France
    \and
    I. Physikalisches Institut der Universit\"at zu K\"oln, Z\"ulpicher Stra{\ss}e 77, 50937 K\"oln, Germany
    \and
    Department of Physical Sciences, The Open University, Milton Keynes MK7 6AA, UK
    \and
   Instituto de Ciencia de Materiales de Madrid, CSIC, Sor Juana In\'es de la Cruz, 3,
28049 Madrid, Spain.
    \and
    Universit\'e de Toulouse, UPS-OMP, IRAP, 31400 Toulouse, France
    \and
    CNRS, IRAP, 9 Av. Colonel Roche, BP 44346, 31028 Toulouse Cedex 4, France
    \and
    LERMA, Observatoire de Paris, PSL Research University, Ecole Normale
Sup\'erieure, CNRS,  75014 Paris
	\and
	Sorbonne Universit\'es, UPMC Paris 06, CNRS, LERMA, 75005 Paris
   } 

   \date{Received <date>/ accepted <date>}

 
  \abstract
   {The methylidyne cation (CH$^+$) and hydroxyl (OH) are key molecules in the warm interstellar chemistry, but their formation and excitation mechanisms are not well understood. Their abundance and excitation are predicted to be enhanced by the presence of vibrationally excited H$_2$ or hot gas ($\sim$500$-$1000~K) in photodissociation regions with high incident FUV radiation field. The excitation may also originate in dense gas ($>10^5$~cm$^{-3}$) followed by nonreactive collisions with H$_2$, H, and electrons. Previous observations of the Orion Bar suggest that the rotationally excited CH$^+$ and OH correlate with the excited CO, a tracer of dense and warm gas, and formation pumping contributes to CH$^+$ excitation.}
   {Our goal is to examine the spatial distribution of the rotationally excited CH$^+$ and OH emission lines in the Orion Bar in order to establish their physical origin and main formation and excitation mechanisms.}
   {We present spatially sampled maps of the CH$^+$ J=3-2 transition at 119.8~$\muup$m and the OH $\Lambda$-doublet at 84~$\muup$m in the Orion Bar over an area of 110$\arcsec \times$110$\arcsec$ with \textit{Herschel} (PACS). We compare the spatial distribution of these molecules with those of their chemical precursors, C$^+$, O and H$_2$, and tracers of warm and dense gas (high-J CO). We assess the spatial variation of CH$^+$ J=2-1 velocity-resolved line profile at 1669~GHz with \textit{Herschel} HIFI spectrometer observations.}
   {The OH and especially CH$^+$ lines correlate well with the high-J CO emission and delineate the warm and dense molecular region at the edge of the Bar. While notably similar, the differences in the CH$^+$ and OH morphologies indicate that CH$^+$ formation and excitation are strongly related to the observed vibrationally excited H$_2$. This, together with the observed broad CH$^+$ line widths, indicates that formation pumping contributes to the excitation of this reactive molecular ion. Interestingly, the peak of the rotationally excited OH 84~$\muup$m emission coincides with a bright young object, proplyd 244-440, which shows that OH can be an excellent tracer of UV-irradiated dense gas.}
   {The spatial distribution of CH$^+$ and OH revealed in our maps is consistent with previous modeling studies. Both formation pumping and nonreactive collisions in a UV-irradiated dense gas are important CH$^+$ J=3-2 excitation processes. The excitation of the OH $\Lambda$-doublet at 84~$\muup$m is mainly sensitive to the temperature and density.}


   \maketitle
%

\section{Introduction}
\label{sect:intro}

The methylidyne cation (CH$^+$) and hydroxyl (OH) have been observed in different environments from the local \citep[e.g.,][]{Storey1981, Nagy2013, Dawson2014} to the extragalactic ISM \citep[e.g.,][]{Schmelz1986, Baan1992, Darling2002, Spinoglio2012, Rangwala2014}. They are key molecules in the warm interstellar chemistry, and because they require ultraviolet (UV) radiation and high temperatures to form, they trace specific physical processes in the ISM \citep[e.g., ][]{Gerin2016}. Thus, it is expected that in addition to the physical conditions of the regions they are observed in, the spatial distribution of the rotationally excited far infrared (FIR) emission of these species might give clues on their formation and excitation processes. \textit{Herschel} allows, for the first time, to study the spatial distribution of these lines. In this paper, we present the first spatially resolved maps of these lines in the Orion Bar.

Atoms and hydrogen molecules and radicals are the first chemical building blocks of the ISM. In warm gas CH$^+$ is believed to form mainly as a product of the $C^+ + H_2$ reaction. This formation route has a very high endothermicity of 0.374~eV (4300~K), and it has been suggested that this barrier could be overcome by reactive collisions with the vibrationally excited H$_2$ in strongly irradiated PDRs \citep[e.g.,][]{White1984, Lambert1986, Jones1986, Agundez2010, Naylor2010, Godard2013, Nagy2013, Zanchet2013}. In diffuse interstellar clouds with low far-ultraviolet (FUV) radiation field and very low density, shocks and turbulence \citep[e.g.,][]{Elitzur1978, PineaudesForets1986, Godard2009, Godard2012, Falgarone2010a, Falgarone2010b} have been proposed to overcome the high endothermicity. In addition, \citet{Lim1999} found that the observed abundances of CH$^+$ by \citet{Cernicharo1997} are explained through thermal reaction between C$^+$ and H$_2 \,(v=0)$ if the gas is hot enough ($>$1000~K). 

The reaction $C^+ + H_2 (\rm v)$ becomes exothermic when H$_2$ is in vibrationally excited states \citep{Godard2013, Zanchet2013}. On the other hand, CH$^+$ is highly reactive and easily destroyed by reactive collisions with H$_2$, H, and electrons. \citet{Godard2013} and \citet{Zanchet2013} find that with the chemical destruction rates considered typical for PDR environments, all levels of CH$^+$ ($J\ge2$) are very sensitive to the formation pumping since the time-scale for the chemical reaction becomes comparable or shorter than that of the nonreactive collision. Collisional excitation with H$_2$, H, and electrons could, however, be important for the lowest rotational transitions. In contrast, radiative pumping is predicted to have only a marginal effect even in a strong FIR radiation field.

The OH radical is a key intermediary molecule in forming other important PDR tracers like H$_2$O, CO$^+$, O$_2$, NO and SO$^+$ in the ISM. On the surface of high FUV-flux PDRs the OH formation is expected to be dominated by the endothermic reaction with H$_2$ and atomic oxygen O$^0$ \citep{Goicoechea2011, Hollenbach2012}. The route is endothermic by  0.08~eV ($\sim$900~K) and with an activation barrier of 0.4~eV ($\sim$4800~K). Similarly to CH$^+$, the possibility of the FUV-pumped vibrationally excited H$_2$ enhancing the abundance and excitation of OH has been suggested.

\citet{Goicoechea2011} made the first detection of rotationally excited CH$^+$ and OH emission lines with \textit{Herschel}/PACS towards the \textit{CO$^+$ peak} \citep{Storzer1995} at the edge of the Orion Bar PDR, one of the nearest, nearly edge-on luminous PDRs (with a FUV radiation field of a few 10$^4$ in Draine units). While their pointed observations cover a small area (47$\arcsec \times$~47$\arcsec$), they hint at a possible spatial correlation between these lines and suggest that the rotational OH emission originates in small irradiated dense structures ($n_H\sim$10$^{6-7}$~cm$^{-3}$ and T$_k\sim$160$-$220~K). The Orion Bar is thought to contain different physical structures with an interclump medium at medium density ($n_H\sim 10^4-10^{5}$~cm$^{-3}$) and high density clumps ($n_H\sim$10$^{6-7}$~cm$^{-3}$) \citep[e.g., ][]{Tielens1993, Tauber1994, Lis2003, Lee2013}. A recent analysis of the high-J CO, H$_2$, OH, and CH$^+$ emission lines using the PDR Meudon code by Joblin et al. (in prep.) shows that these lines are very sensitive to the thermal pressure and suggests $P\sim2 \times 10^8$~K~cm$^{-3}$ for the emitting structures. However, this study uses pointed observations and in order to understand the CH$^+$ and OH formation and excitation in detail, we need to understand their spatial distribution and how they compare with other tracers.
 
Using high spectral resolution \textit{Herschel}/HIFI observations, \citet{Nagy2013} find that the CH$^+$ J=1-0 and J=2-1 lines have line widths of $\Delta V \sim 5$~km~s$^{-1}$ towards the Orion Bar, larger than those of other molecular lines, which typically show line widths of 2$-$3~km~s$^{-1}$. This line broadening could result from formation pumping leading to a non-thermalized velocity distribution.

In this paper, we study the spatial distribution of the rotationally excited CH$^+$ and OH emission lines using fully sampled PACS maps of CH$^+$ transition J=3-2 at 119.8~$\muup$m and OH 84~$\muup$m $\Lambda$-doublet towards the Orion Bar. These are the first fully sampled maps of these emission lines in a PDR. Observational constraints on the excited CH$^+$ and OH spatial distribution and comparison with tracers of warm and dense gas and vibrational excited H$_2$ are needed to establish their physical origin and their main excitation mechanisms (collisions versus pumping). Given the similar critical densities ($\sim$10$^{10}$~cm$^{-3}$) and upper level energies (E/k$\sim$250~K) of the targeted lines, our observations allow us to study how the chemical reaction with H$_2 \, (v>0)$ affects the formation and excitation of CH$^+$ and OH.

We first describe the \textit{Herschel} observations and data reduction of the CH$^+$ and OH lines, as well as the lines observed for comparison, in Sect. \ref{sect:observations}. We discuss the line detection, spatial morphology and origin of the CH$^+$ J=3-2 and OH 84~$\muup$m lines in Sect. \ref{sect:morphology}. In Sect. \ref{sect:H2_comparison} we investigate the CH$^+$ and OH formation and chemical pumping excitation mechanism via reaction with vibrationally excited H$_2$. In this section we also compare OH with H$_2$O, since OH can also be the product of H$_2$O photodissociation in the gas unshielded against FUV radiation. In Sect. \ref{sect:HIFI}, we discuss the CH$^+$ velocity dispersion. The detection of a proplyd in the OH map is described in Sect. \ref{sect:proplyd}. We conclude and summarize the findings of the paper in Sect. \ref{sect:conclusions}.


\section{Observations and data reduction}
\label{sect:observations}

\subsection{CH$^+$ J=3-2 and OH lines observed with PACS}

We observed the Orion Bar over an area of 110$\arcsec \times$110$\arcsec$ with the Photoconductor Array Camera and Spectrometer \citep[PACS,][]{Poglitsch2010} onboard \textit{Herschel} Space Observatory. In particular, we mapped the following lines: CH$^+$ J=3-2 (120~$\muup$m), OH 84~$\muup$m $\Lambda$-doublet (84.4 and 84.6~$\muup$m, transitions $^2\Pi_{3/2}$ J=7/2$^+$-5/2$^-$ and $^2\Pi_{3/2}$ J=7/2$^-$-5/2$^+$, respectively), and OH 119 $\muup$m $\Lambda$-doublet (119.2 and 119.4~$\muup$m, transitions $^2\Pi_{3/2}$ J=5/2$^-$-3/2$^+$ and $^2\Pi_{3/2}$ J=5/2$^+$-3/2$^-$, respectively). Table \ref{table:properties} lists the observed transitions, their wavelengths, critical densities, upper level energy temperatures, and Einstein coefficients.
\begin{table*}
 \caption[properties]{Properties of the observed lines: wavelength ($\lambda$), transition, critical density (n$_{crit}$), upper level energy temperature (E$_u$/k), and Einstein coefficient (A$_{ij}$). Numbers in parenthesis are power of 10. The critical densities have been calculated using the collisional rates with $a$) ortho-H$_2$ \citep{Offer1994}, $b$) He \citep{Hammami2008,Hammami2009}, $c$) H$_2$ \citep{Yang2010},  $d$) H$^0$ \citep{Launay1977}, and $e$) para-H$_2$ \citep{Dubernet2002, Phillips1996}.}
 	\vspace{0,2cm}
 	\centering
 	\begin{tabular}{ lccccc }
   	\hline
   	\hline
   	 & $\lambda$ [$\muup$m] & transition & n$_{crit}$ [cm$^{-3}$] & E$_u$/k [K]  & A$_{ij}$ [s$^{-1}$] \\
   	\hline 
   	OH        & 84.6  & $^2\Pi_{3/2}$ J=7/2$^-$-5/2$^+$ & 1 (10)$^a$ & 291  & 5.20 (-01) \\ 
   	OH        & 119.4 & $^2\Pi_{3/2}$ J=5/2$^+$-3/2$^-$ & 1 (09)$^a$ & 120  & 1.38 (-01) \\
   	CH$^+$    & 119.8 & J=3-2                           & 4 (09)$^b$ & 240  & 2.20 (-01) \\
   	CH$^+$    & 179.6 & J=2-1                           & 6 (08)$^b$ & 120  & 6.10 (-02) \\
   	CH$^+$    & 359.0 & J=1-0                           & 7 (07)$^b$ & 40   & 6.36 (-03) \\
   	CO        & 137.3 & J=19-18                         & 5 (06)$^c$ & 1050 & 6.65 (-04) \\
   	$^{13}$CO & 227.1 & J=12-11                         & 2 (06)$^c$ & 412  & 1.52 (-04) \\
   	C$^{18}$O & 341.5 & J=8-7                           & 6 (05)$^c$ & 190  & 4.47 (-05) \\
   	$[$CII]   & 157.7 & $^2$P$_{3/2}$-$^2$P$_{1/2}$     & 3 (03)$^d$ & 91   & 2.29 (-06) \\
    $[$OI]    & 145.5 & $^3$P$_0$-$^3$P$_1$             & 2 (05)$^d$ & 327  & 1.66 (-05) \\
   	H$_2$O    & 269.3 & 1$_{11}$-0$_{00}$               & 1 (09)$^e$ & 53   & 1.84 (-02) \\
   	H$_2$O    & 398.7 & 2$_{11}$-2$_{00}$               & 3 (08)$^e$ & 137  & 7.05 (-03) \\
   	\hline 
   	\end{tabular}
   	\label{table:properties}
 \end{table*}

\begin{figure}
    \centering
    \includegraphics[scale = 0.6]{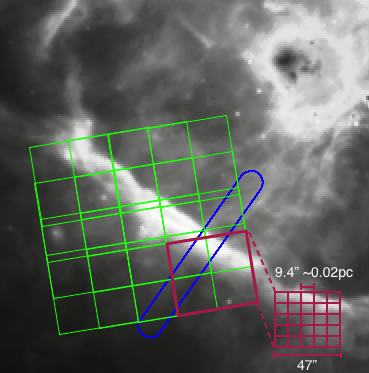}
    \caption[configuration]{Overlay of the PACS observations and HIFI observations on a Spitzer IRAC 8 $\muup$m image of the Orion Bar. A 4 $\times$ 4 raster map of CH$^+$ J=3-2 – 16 overlapping footprints — is shown in green and an example of a footprint is shown in red. The strip of CH$^+$ J=2-1 is shown in blue.}
    \label{pic:config}
\end{figure}

The lines were observed using the unchopped mode and due to the instantaneous wavelength window covered, the CH$^+$ \mbox{J=3-2} observations at 119.8~$\muup$m also contain the OH 119~$\muup$m doublet. The observations were carried out on September 14 and 15, 2012  and the total observation time was 21~009s and 13~795s for OH and CH$^+$, respectively. The observations consist of a fully Nyquist-sampled raster map of 4$\times$4 footprints of the  CH$^+$ \mbox{J=3-2} line, and a fully Nyquist-sampled 5$\times$5 raster map of the OH 84 $\muup$m line. These footprints are composed of 5$\times$5 spatial pixels (spaxels). For each spaxel the line is observed in 16 different spectral scans, each with an up and down scan. The configuration is shown in Fig. \ref{pic:config}, where the raster map is overlaid on top of the 8~$\muup$m IRAC image of the Orion Bar. 
Due to the overlapping footprints, the middle of the map is better sampled and has (slightly) higher S/N than the edges of the map where we only have the data from one footprint. For CH$^+$ \mbox{J=3-2} and OH 84~$\muup$m the typical line detection level outside the Bar is $\sim$10$-$15$\sigma$ and in the Bar $\sim$15$-$30$\sigma$. For OH 119~$\muup$m the typical detection level outside the Bar is $\sim$50$\sigma$ and in the Bar $>$100$\sigma$.

The data were processed using the version 10.0.2843 of the reduction and analysis package HIPE. Since the PACS instrument has a spectral resolution of $\sim$135~km~s$^{-1}$ at 85~$\muup$m and $\sim$300~km~s$^{-1}$ at 120~$\muup$m, the line widths and the hyperfine structures of the OH $\Lambda$-doublets are not resolved in our observations. The spatial resolution is 9$\arcsec$ for the CH$^+$ 120~$\muup$m line and the OH 119~$\muup$m $\Lambda$-doublets, and 6$\arcsec$ for the OH 84~$\muup$m $\Lambda$-doublets.

\begin{figure}
	\centering
	\includegraphics[scale = 0.25]{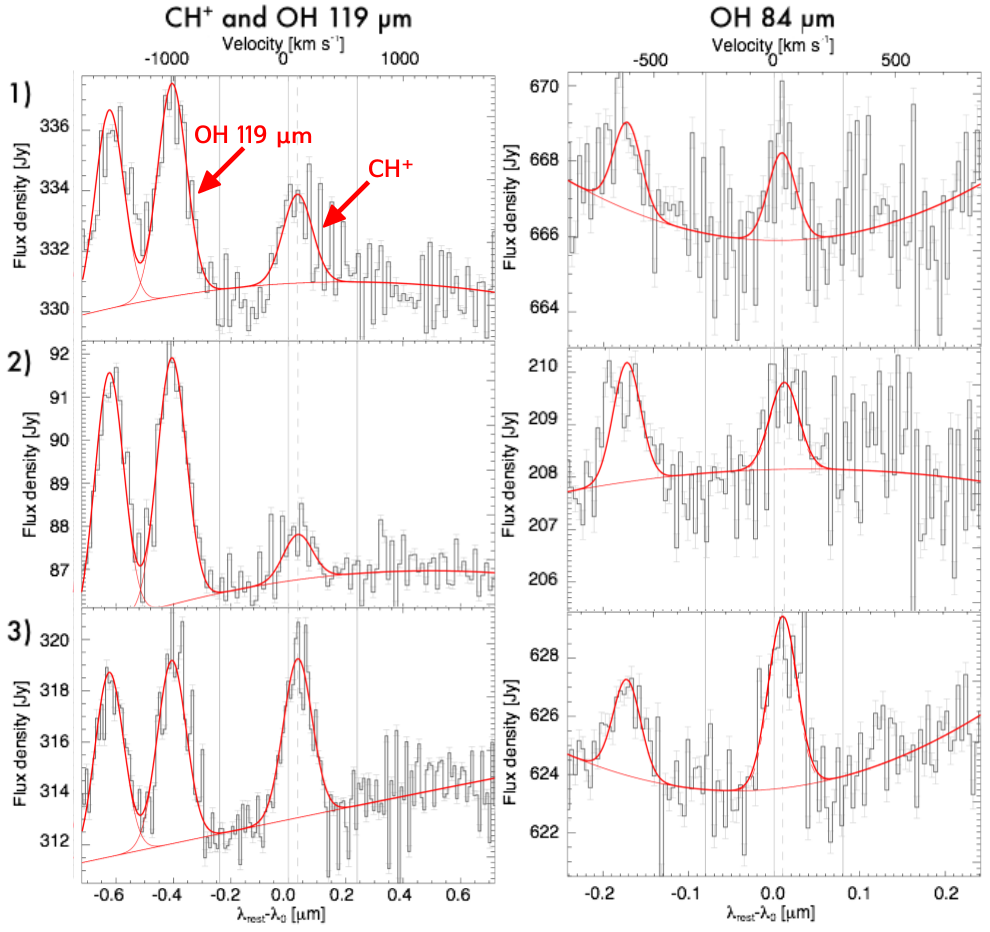}
	\caption[spectra]{Spectra of CH$^+$ J=3-2 (120~$\muup$m) and OH 119~$\muup$m (left) and OH 84~$\muup$m (right) at three positions indicated in panel a) of Fig. \ref{pic:basic_maps}.}
    \label{pic:spectra}
\end{figure}

The line fitting was performed using IDL-based software PACSman version 3.55 \citep{Lebouteiller2012}. Fig. \ref{pic:spectra} gives an example of the observations and line fitting routine. 
Using a polynomial baseline, the lines were fitted with a Gaussian profile and the corresponding line flux integrated. PACSman measures the line fluxes for each spatial pixel independently. To produce the final map, it creates an oversampled pixelated grid of the observations with a 3$\arcsec$ pixel resolution and calculates the average fractional contribution of the given spatial pixels to the relevant position. PACSman also calculates the statistical uncertainties, including the dispersion in the reduction process and the rms of the fit. In our observations, these uncertainties are small and usually amount to at most 5$-$20~\% for the CH$^+$ and OH lines. The relative accuracy between spatial pixels given in the manual is 10~\%\footnote{From the PACS spectroscopy performance and calibration manual. This can be found at \url{http://herschel.esac.esa.int/twiki/bin/view/Public/PacsCalibrationWeb}}. We assume a conservative total uncertainty of 22~\% for the integrated line intensities in the fainter regions, that is the upper limit of a combination of the calibration (10~\%) and line fitting uncertainties (5$-$20~\%). For the Bar, the uncertainty is less than 11~\%.

We also observed the CO \mbox{J=19-18} line at 137.3~$\muup$m with PACS (Parikka et al., in prep.). The observations consist of a fully Nyquist-sampled raster map of 4$\times$4 footprints. For CO \mbox{J=19-18}, the typical line detection level outside the Bar is $\sim$10$\sigma$ and in the Bar $\sim$100$\sigma$. The uncertainty for the integrated line intensities is less than 22~\% with less than 11~\% in the Bar. The spatial resolution is 10$\arcsec$.

\subsection{HIFI and SPIRE observations}

In order to obtain information about the velocity structure of the CH$^+$ lines, we used the Heterodyne Instrument for the Far Infrared \citep[HIFI,][]{DeGraauw2010} to observe the CH$^+$ \mbox{J=2-1} line at 180~$\muup$m along a cut perpendicular to the Orion Bar (see Fig. \ref{pic:config}). The cut is centered at the position $\alpha_{J2000}$= 05$^{\rm h}$35$^{\rm min}$20.61$^{\rm s}$, $\delta_{J2000}$= \mbox{-05$\degr$25$\arcmin$14$\arcsec$} and extends across 2$\arcmin$ with an inclination of PA=52$\degr$. The data were obtained with the High Resolution Spectrometer (HRS). We used the on-the-fly observing mode using half-beam sampling and the nominal HIFI beam size at this frequency (12.3$\arcsec$).  After porting the FITS data into GILDAS format\footnote{Institut de Radioastronomie Millim\'etrique (IRAM): \url{http://iram.fr/IRAMFR/GILDAS/}}, the data processing consisted of a scan-by-scan linear baseline subtraction and resampling to a common frequency grid with a resolution of 0.3~km~s$^{-1}$. The intensity scale is antenna temperature ($T_A^*$), and the typical rms is 0.12~K calculated at a velocity resolution of 0.3~km~s$^{-1}$. The offset (from the central position of $\alpha_{J2000}$= 05$^{\rm h}$35$^{\rm min}$20.61$^{\rm s}$, $\delta_{J2000}$= -05$\degr$25$\arcmin$14$\arcsec$) of the spectra, the calculated area, the velocity ($V_{LSR}$), FWHM line width ($\Delta V$) and the peak temperature [K] are listed in Table~\ref{table:HIFI}.
\begin{table*}
	\caption[CHp J=2-1]{Results from the fitting of the Herschel/HIFI observations of the  CH$^+$ J=2-1 line. The first column lists the offset from the central position of $\alpha_{J2000}$: 05$^{\rm h}$35$^{\rm min}$20.61$^{\rm s}$, $\delta_{J2000}$: -05$\degr$25$\arcmin$14$\arcsec$, and subsequent columns give the area [K~km~s$^{-1}$], velocity $\Delta V$ [km~s$^{-1}$], and the peak temperature [K]. The uncertainties are given in parenthesis. For T$_{peak}$ the uncertainty is approximately 10 \% (calibration). The rest frequency for CH$^+$ J=2-1 line is 1669.2813 GHz.}
	\vspace{0,2cm}
	\centering
	\begin{tabular}{ lcccc }
  		\hline
  		\hline
    	Offset & Area [K~km~s$^{-1}$]  & $V_{LSR}$ [km~s$^{-1}$] & $\Delta V$ [km~s$^{-1}$] & $T_{peak}$ [K] \\
  		\hline
		$[$-26$\arcsec$, 40$\arcsec$] & 2.5 (0.2) & 8.7 (0.1)  & 3.4 (0.3) & 0.69 \\
		$[$-13$\arcsec$, 20$\arcsec$] & 3.2 (0.2) & 9.6 (0.1)  & 3.4 (0.2) & 0.88 \\
		$[$0$\arcsec$, 0$\arcsec$]	  & 6.8 (0.2) & 10.6 (0.1) & 4.0 (0.2) & 1.57 \\
		$[$13$\arcsec$, -20$\arcsec$] & 2.0 (0.2) & 10.9 (0.1) & 3.4 (0.3) & 0.55 \\
		$[$26$\arcsec$, -40$\arcsec$] & 0.9 (0.2) & 11.1 (0.3) & 3.1 (0.7) & 0.28 \\
  		\hline 
  	\end{tabular}
  	\label{table:HIFI}
\end{table*}
\begin{figure*}
	\centering
	\includegraphics[scale = 0.34]{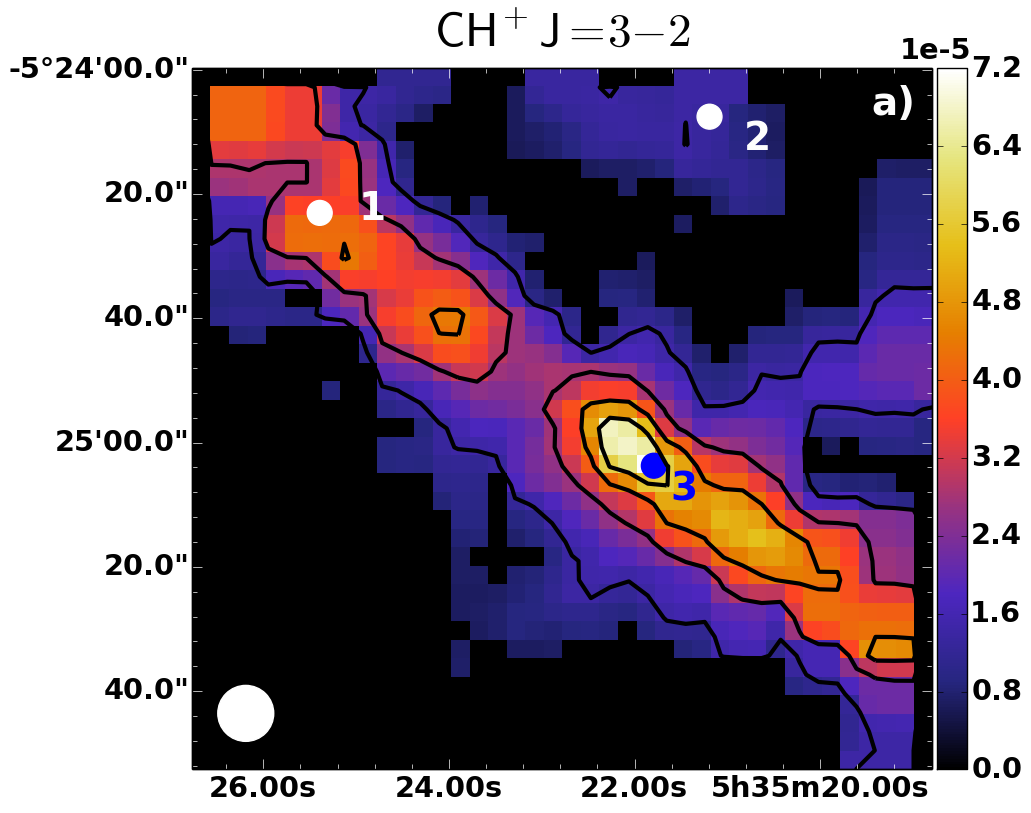} \hfill
	\includegraphics[scale = 0.34]{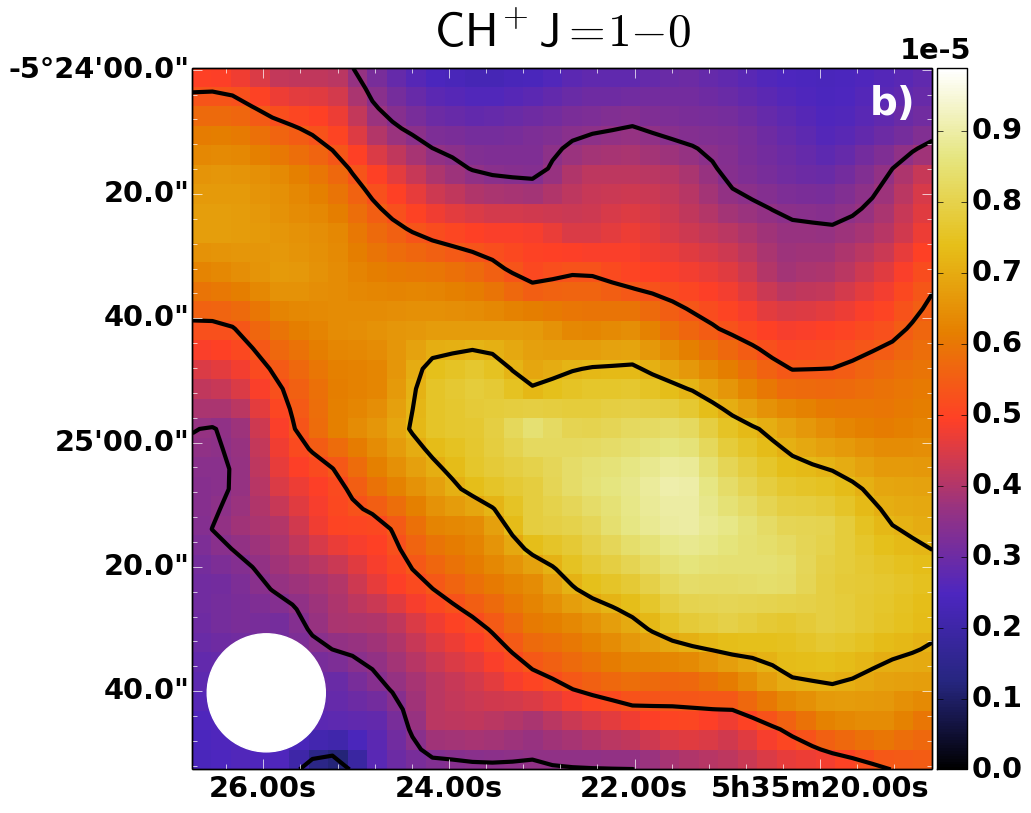}
	\includegraphics[scale = 0.34]{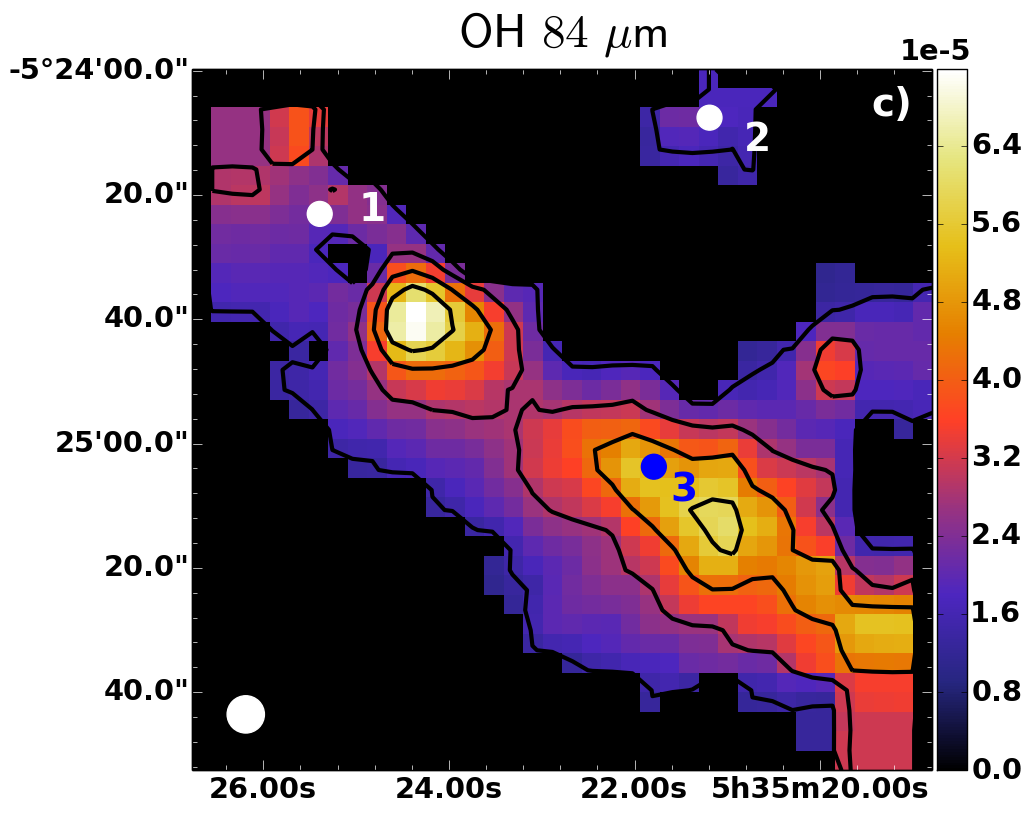}\hfill
	\includegraphics[scale = 0.34]{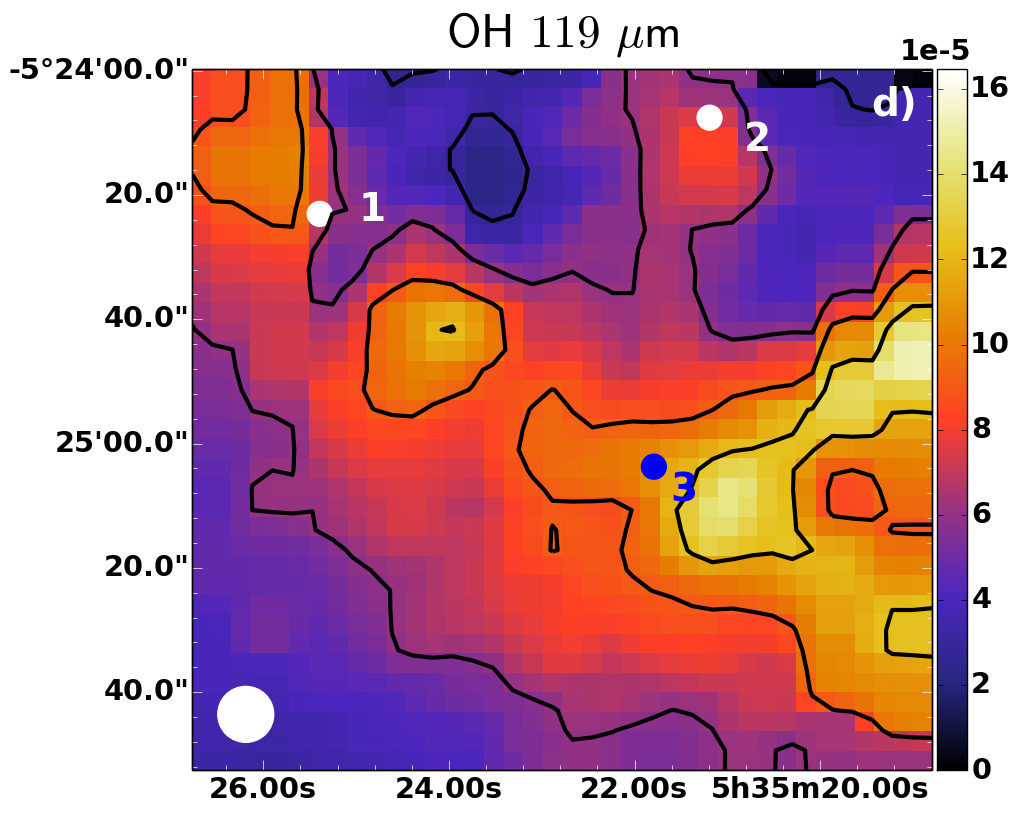}	
	\caption[morphology]{Morphology of different lines: a) CH$^+$ J=3-2, b) CH$^+$ J=1-0, c) OH 84.6~$\muup$m, and d) OH 119.4 $\muup$m. Contours with contour steps of 20~\%  of the peak emission in the Bar. The intensities are in units of erg~s$^{-1}$~cm$^{-2}$~sr$^{-1}$. The black pixels are pixels where lines are not detected. The beam size (6$\arcsec$ for OH 84 $\muup$m, 9$\arcsec$ for CH$^+$ J=3-2 and OH 119~$\muup$m, and 19$\arcsec$ for CH$^+$ J=1-0) is shown in the bottom left corner. The positions where the spectra in Fig. \ref{pic:spectra} have been taken from are also marked in the pictures with corresponding numbers.}
    \label{pic:basic_maps}
\end{figure*}

We have also observed CH$^+$ J=1-0 at 359~$\muup$m and water lines 1$_{11}$-0$_{00}$ at 269~$\muup$m and 2$_{11}$-2$_{02}$ at 399~$\muup$m with the Spectral and Photometric Imaging Receiver \citep[SPIRE,][]{Griffin2010}. The observations were taken on September 20, 2010 in the high-resolution SPIRE FTS full-sampling mode. We include the water lines in our observations as OH can also form via the photodissociation of H$_2$O. Within the SPIRE range of these observations, a wealth of rotational lines of CO and its isotopologues were detected \citep[][Parikka et al., in prep.]{Habart2010}. The $^{13}$CO J=12-11 line at 227~$\muup$m and C$^{18}$O J=8-7 line at 342~$\muup$m maps are presented.

The SPIRE data were processed using HIPE 11.0.1. To achieve a better angular resolution, the super-resolution method SUPREME was applied to the FTS data as in \citet{Kohler2014}. In Figs. \ref{pic:basic_maps} and \ref{pic:water_comparison} we give a FWHM using a Gaussian fit with the same bandwidth of the equivalent beam. It should, however, be noted that the Point Spread Function (PSF) is non-Gaussian and a FWHM does not account for the shape of the beam. The gain in resolution with the SUPREME method depends on the wavelength. The resolution improves, e.g., from 32.6$\arcsec$ to 19.0$\arcsec$ at 400~$\muup$m or from 40.1$\arcsec$ to 25.8$\arcsec$ at 650~$\muup$m. After SUPREME processing, the spatial resolution for the CH$^+$ J=1-0, H$_2$O 399 $\muup$m and C$^{18}$O J=8-7 lines  is  between 17$\arcsec$ and 19$\arcsec$, and for the H$_2$O 269 $\muup$m and $^{13}$CO J=12-11 line between 13$\arcsec$ and 14$\arcsec$. For a detailed description of the reduction procedure method, fitting routines, and PSF see the SUPREME web site\footnote{\url{http://www.ias.u-psud.fr/supreme/hipeplugin.php}, see also a draft available at: \url{http://hal.archives-ouvertes.fr/hal-00765929}}. The lines were fitted with the HIPE Spectrometer Cube fitting procedure. We assume a total uncertainty of 36~\% for the integrated line intensities, which includes the calibration and line fitting uncertainties.


\section{Line emission detection and spatial morphology}
\label{sect:morphology}

\subsection{Origin of the CH$^+$ emission}

In this section we present the observations of CH$^+$ \mbox{J=3-2} transition at 120~$\muup$m and J=1-0 transition at 359~$\muup$m. Maps of the integrated line intensities are shown in Fig. \ref{pic:basic_maps}. We show line spectra for CH$^+$ \mbox{J=3-2} at three positions specified in the maps (2 in the Bar, 1 in front, Fig. \ref{pic:spectra}).

Fig. \ref{pic:basic_maps} shows that the CH$^+$ J=3-2 line emission is well detected in the Bar, but we do not detect it behind the Bar. The emission  delineates the edge of the Bar, which indicates that it is a good tracer of the warm molecular zone and very sensitive to the physical conditions prevailing in the Bar. The CH$^+$ \mbox{J=1-0} emission is much broader and originates both from the Bar and from the face-on Orion Molecular cloud surface in the background. The J=1-0 transition is also more easily excited which could affect the breadth of the emission. Emission from the UV-illuminated surface of Orion molecular cloud complex behind the Bar has been seen in several tracers, such as C$^+$ \citep{Bernard-Salas2012, Goicoechea2015a}. The CH$^+$ J=3-2 is less extended than the J=1-0 line, and may originate from higher {density} structures in the Bar. In our maps, the CH$^+$ J=3-2 line emission not associated with the Bar comes from the molecular cloud complex in the Northern part of the map and in the Western region of the map (Orion Ridge). Emission in these regions have also been observed in [CII] and [OI] \citep{Bernard-Salas2012}. 

\begin{figure}
	\centering
	\includegraphics[scale = 0.34]{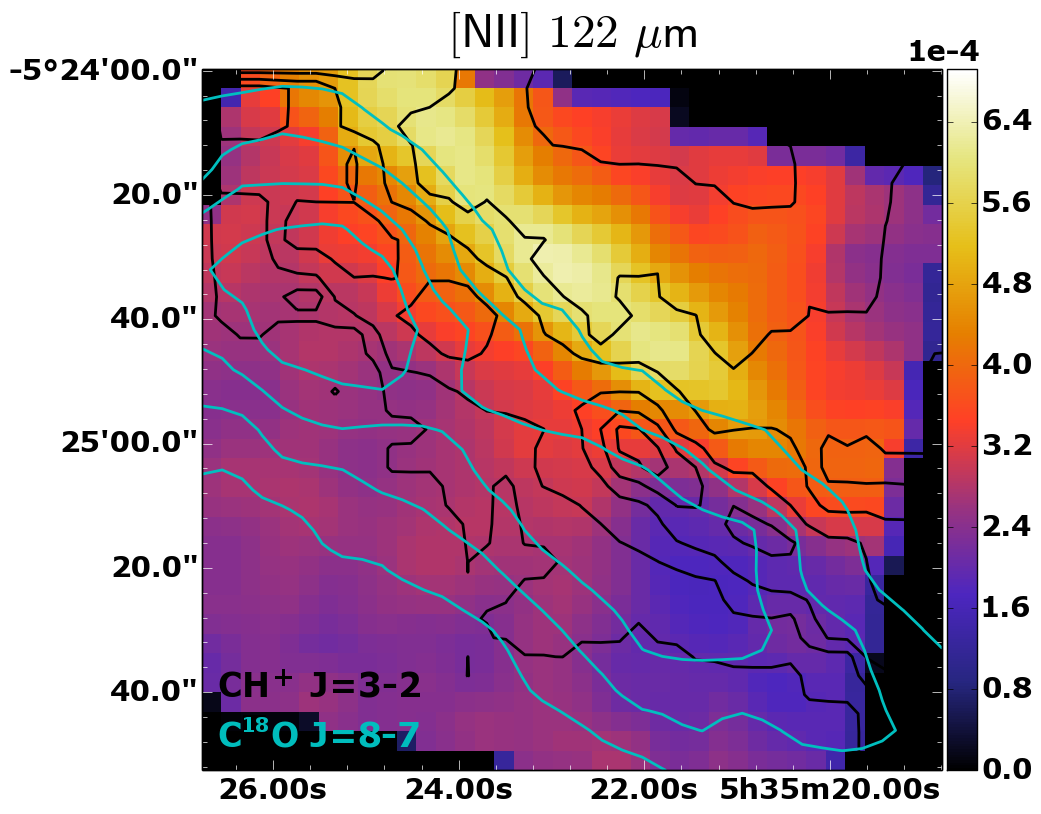}
	\caption[IF]{[NII] 122~$\muup$m map with contours of CH$^+$ J=3-2 (black) and C$^{18}$O J=8-7 (cyan, peak emission 5.6$\times 10^{-6}$). Contours are with contour steps of 20~\%  of the peak emission in the Bar. The intensities are in units of erg~s$^{-1}$~cm$^{-2}$~sr$^{-1}$.}
    \label{pic:Np_CHp_C18O}
\end{figure}

Fig. \ref{pic:Np_CHp_C18O} shows that the warmest PDR molecular lies between the [NII] 122~$\muup$m emission tracing the hot ionization front\footnote{[NII] 122~$\muup$m peaks in the same place where other ionization front tracers peak, e.g. OI 1.32~$\muup$m \citep{Lis2003}.} \citep{Bernard-Salas2012}, and the C$^{18}$O J=8-7 line which traces colder gas.
Most of the molecules are photodissociated in the ionized layer in front of the Bar. Behind the Bar and further away, the conditions (low FUV flux, low temperature) are not favorable to form and/or excite CH$^+$. 

\begin{figure*}
	\centering
	\includegraphics[scale = 0.23]{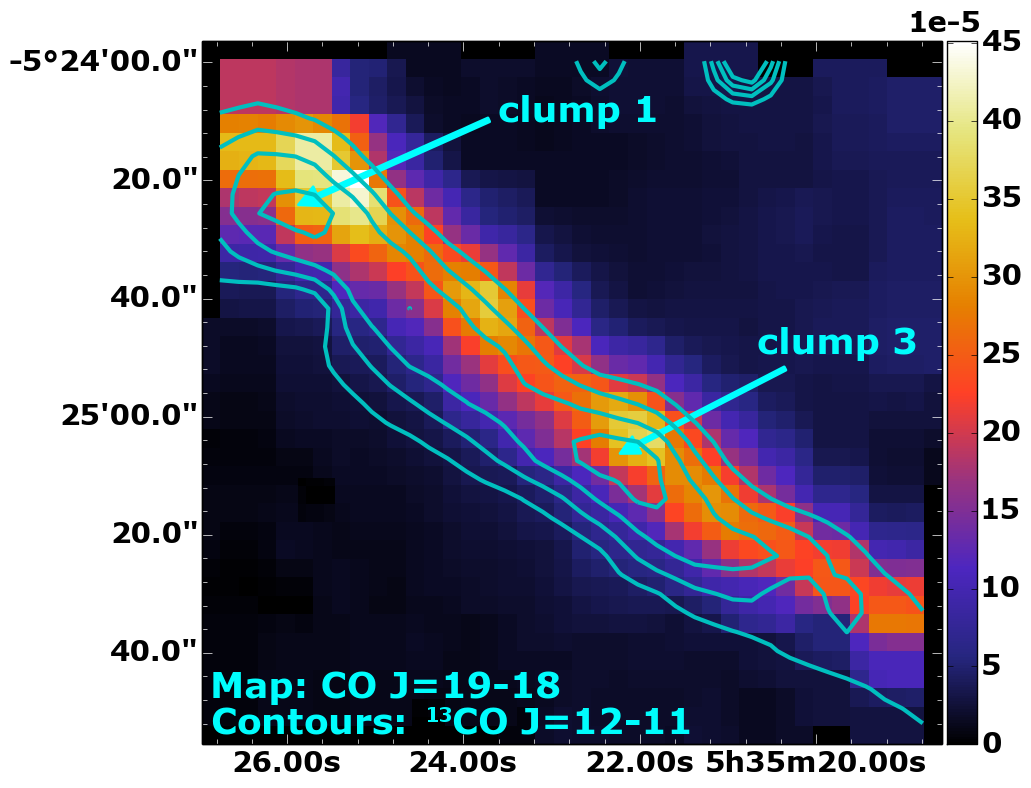} \hfill
	\includegraphics[scale = 0.23]{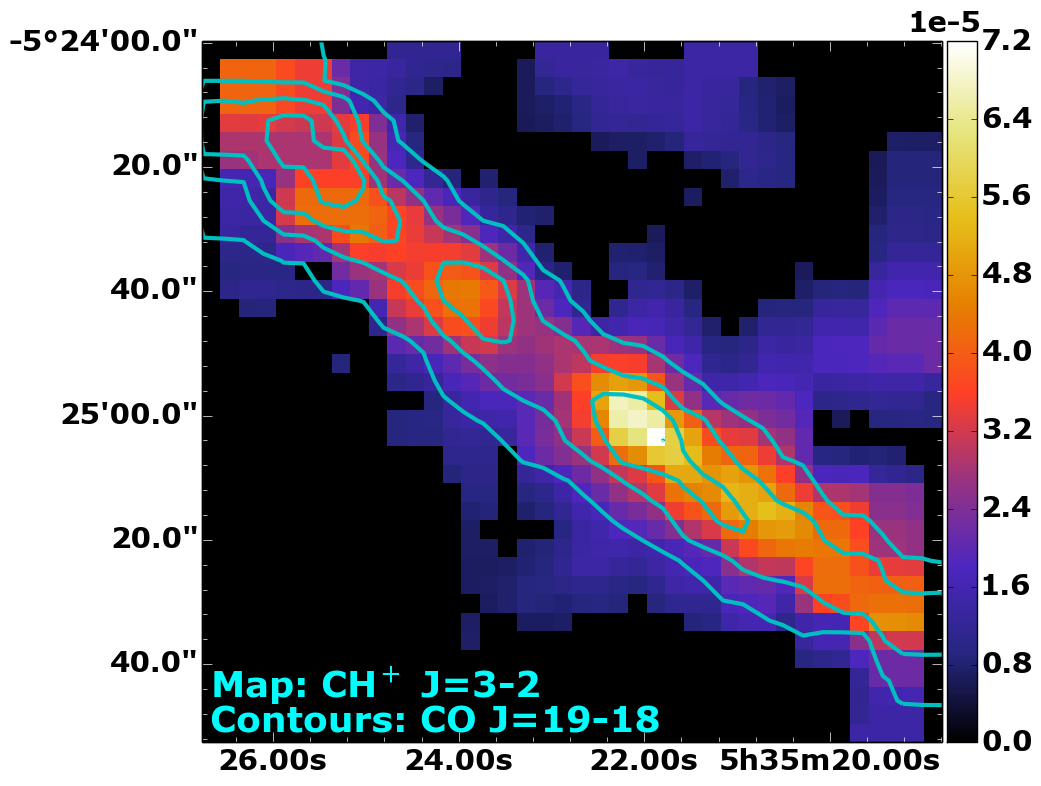} \hfill
	\includegraphics[scale = 0.23]{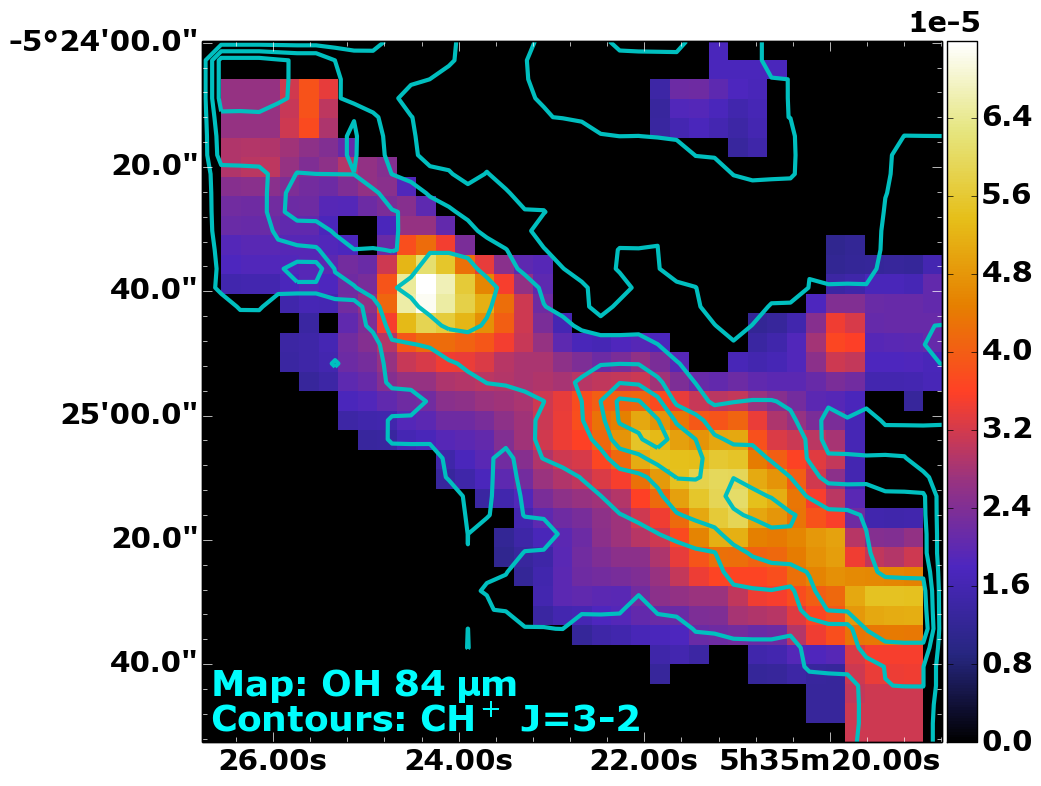} 
	\caption[co]{Map of CO J=19-18 (137~$\muup$m) integrated intensity with $^{13}$CO J=12-11 contours (left). Map of CH$^+$ J=3-2 integrated intensity with CO J=19-18 contours (middle) and OH 84.6 $\muup$m with CH$^+$ J=3-2 contours (right). All contours are with contour steps of 20~\%  of the peak emission in the Bar starting from 10~\%, except $^{13}$CO starting from 30~\%. The intensities are in units of erg~s$^{-1}$~cm$^{-2}$~sr$^{-1}$. The maps are not convolved as the resolutions are similar. Clumps 1 and 3 \citep{Lis2003} are marked in the figure on the left.}
    \label{pic:CO_comparison}
\end{figure*}

Fig. \ref{pic:CO_comparison} shows the comparison of CO J=19-18 map with $^{13}$CO J=12-11, a tracer of the warm and dense phase.
Both of these lines show clumps which correspond to the ones detected from the ground in H$^{13}$CN \citep{Lis2003} and in \mbox{CS J=2-1} \citep{Lee2013}. We can clearly see the clumps\footnote{We have adopted the numbering of \citet{Lis2003}.} 1 and 3, respectively at the North and South part of the Bar. The clumps have diameters of about $\simeq$9$\arcsec$ or 0.02~pc and are marginally resolved by \textit{Herschel}. The clumps seen in $^{13}$CO J=12-11 correspond exactly to the ones seen in H$^{13}$CN and \mbox{CS J=2-1}. The CO J=19-18 map shows the same clumps, but shifted by $\sim$5$\arcsec$ toward the PDR front. The CO peaks could, thus, be associated to the surface of the high density zones in the Bar. Fig. \ref{pic:CO_comparison} also compares CH$^+$ J=3-2 with CO J=19-18 and OH 84~$\muup$m with CH$^+$ J=3-2. The clumps can be seen in CH$^+$, while in OH we see only clump 3. The intensity peak of CH$^+$ J=3-2 (and J=1-0) coincides with the clump 3, while CO J=19-18 (and C$^{18}$O J=8-7) peaks at clump 1. Clump 1 has a larger column density, suggesting that the optically thin high-J CO lines are more sensitive to the total column density of warm gas than CH$^+$. Like CH$^+$, most of the surface tracers, such as H$_2$ v=1-0 S(1) or optically thick lines (e.g., CO J=6-5), peak at the Southern part of the Bar.

\begin{figure*}
	\centering
	\includegraphics[scale = 0.31]{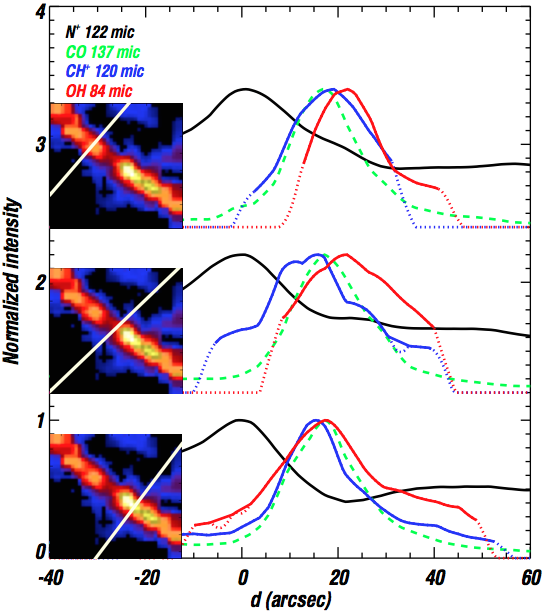}
	\includegraphics[scale = 0.31]{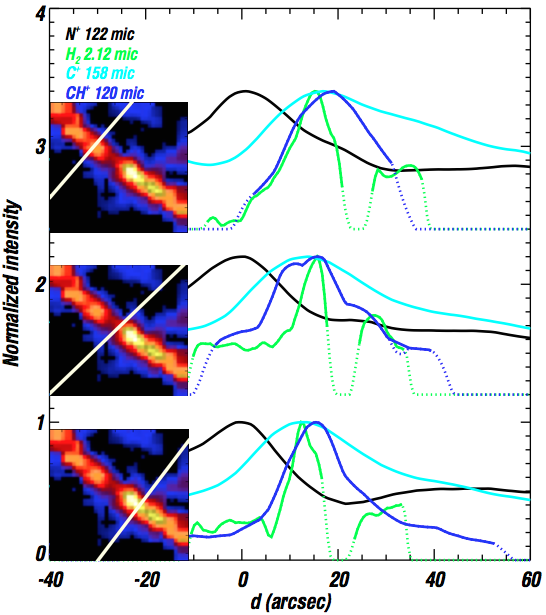}
	\includegraphics[scale = 0.31]{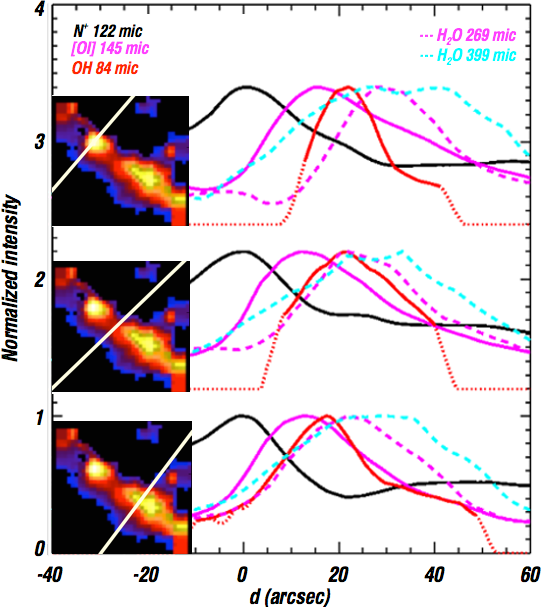}
	\caption[IF]{Profile cuts across the Bar at positions indicated in the small CH$^+$ J=3-2 and OH 84~$\muup$m maps. The intensity of the lines have been normalized to the maximum emission in the Bar and the lines are not convolved to the same beam, so that the structures seen in the unconvolved maps are not lost. The distance is from the ionization front marked by the peak of [NII] emission. The dashed part of the H$_2$ profile in the cuts indicates the slit seen in the map (see Fig. \ref{pic:H2_comparison_maps}) and the positions of the cuts are indicated in the small maps next to each cut.}
    \label{pic:cuts}
\end{figure*}

Fig. \ref{pic:cuts} presents the profile of the emission of various lines along a cut intersecting the Bar at three different positions. 
The profile of the CH$^+$ J=3-2  line emission is resolved with an average FWHM of 20$\arcsec$, which is wider than the PSF (9$\arcsec$). This FWHM is similar to that of the CO J=19-18 line. The observed width is most likely the result of the Bar being tilted to the observer, an effect already inferred in other studies \citep[e.g.,][]{Hogerheijde1995, Allers2005, Pellegrini2009}. Using an inclination angle of $\theta$=7$^{\circ}$ \citep{Hogerheijde1995, Pellegrini2009}, and a length of the PDR along the line of sight of 0.35~pc \citep{Bernard-Salas2012} needed to reproduce the [CII] and [OI] emission, we find that a width of $sin(\theta) \times l_{PDR}\sim$0.04 pc (or $\sim$20$\arcsec$) is in agreement with the observed one. The CH$^+$ and high-J CO emission come from the inclined thin surface of the Bar. 
We stress that higher angular resolution (1$\arcsec$) observations of rotationally and vibrationally excited H$_2$ originating from the thin PDR surface show a smaller width of $\sim$10$\arcsec$ (see Fig. \ref{pic:cuts}) than those observed for CH$^+$ and high-J CO. Thus, while our data shows a connection with the clumps seen in H$^{13}$CO and CS, the CH$^+$ and high-J CO emission may also originate from unresolved structures close to the cloud surface. Indeed, small-scale, high-density HCO$^+$ J=4-3 emission structures have been recently resolved by ALMA \citep{Goicoechea2016}.


\subsection{Origin of the OH emission}
\label{sect:origin_OH}

In this section we present the observations of OH 84~$\muup$m and 119~$\muup$m $\Lambda$-doublets observed with PACS. The two lines of the $\Lambda$-doublets have the same distribution and almost the same intensities, and therefore one line map is presented for each wavelength, namely 84.6 $\muup$m ($^2\Pi_{3/2}$ J=7/2$^-$-5/2$^+$) and 119.4 $\muup$m ($^2\Pi_{3/2}$ J=5/2$^+$-3/2$^-$). The intensities vary only slightly (by $\sim$30~\%, see Fig. \ref{pic:spectra}) between the two $\Lambda$-doublets for each wavelength. Asymmetries are predicted to be small when collisions with ortho-H$_2$ dominate (i.e., in the warm gas) and when FIR radiative pumping plays a role \citep{Offer1992, Offer1994, Goicoechea2011}. 

The first rotationally excited OH lines at 84~$\muup$m are well detected in the Bar and also in the Orion Ridge and Northern region in front of the Bar (see Fig. \ref{pic:basic_maps}). Like CH$^+$ J=3-2 {in Fig. \ref{pic:CO_comparison}}, the OH 84~$\muup$m $\Lambda$-doublet traces the warm molecular layer, but the OH emitting zone is more extended behind the Bar (see also Fig. \ref{pic:cuts}). This could result from the fact that the formation of OH does not depend on H$_2$~v>0 like CH$^+$ (see Sect. \ref{sect:H2_comparison}) and that the excitation of OH is mainly sensitive to the high density as suggested by previous studies. \citet{Goicoechea2011} suggest that the OH 84~$\muup$m emission is coming from unresolved structures with n(H)$\sim$10$^{6-7}$~cm$^{-3}$ and T$_{kin} \sim$160$-$220~K whereas CH$^+$ emission can be explained by a density of n(H)$\sim$10$^5$~cm$^{-3}$ and T$_{kin} \sim$500$-$1000~K \citep[][Joblin et al., in prep]{Nagy2013}. In the OH 84~$\muup$m map, there are two main density enhancements in the Bar, one of which may be associated with a dense structure (referred to as clump 3), and another that coincides with an externally illuminated protoplanetary disk, which will be discussed in Sect. \ref{sect:proplyd}.

The OH ground-state rotational lines at 119~$\muup$m are clearly detected everywhere in the mapped area (panel d in Fig. \ref{pic:basic_maps}). Since the OH 119~$\muup$m $\Lambda$-doublets are more easily excited than the 84~$\muup$m $\Lambda$-doublets, they show a more extended spatial distribution (including the Orion Ridge and the molecular cloud complex in the North). \citet{Goicoechea2011} used nonlocal, non-LTE radiative transfer model including FIR pumping, to fit several FIR OH rotational lines (at $\sim$119~$\muup$m, $\sim$84~$\muup$m, $\sim$163~$\muup$m, $\sim$79~$\muup$m, and $\sim$65~$\muup$m) detected with \textit{Herschel}/PACS towards a small region\footnote{One PACS footprint, 47$\arcsec \times$47$\arcsec$.} in the Orion Bar. The best fit was found for unresolved structures with T$_{kin} \sim$200~K, n(H)$\sim$10$^6$~cm$^{-3}$ and a source-averaged OH column density of N(OH)=10$^{15}$~cm$^{-2}$. For these conditions, the OH 119~$\muup$m $\Lambda$-doublets are optically thick ($\tau>$10), but the OH 84~$\muup$m $\Lambda$-doublets are optically thin (or effectively optically thin, $\tau<$5 and $T_{\rm ex}\ll T_{\rm kin}$). Therefore, OH 84~$\muup$m line intensities are expected to be proportional to N(OH) in the Orion Bar. 

In conclusion, the OH 84~$\muup$m and  CH$^+$ J=3-2 line emissions correlate well, albeit with some dispersion reflecting the difference in emission discussed above. The observed general correlation was expected given the similar critical densities and upper level energies and a main formation route with H$_2$ for both lines. However, owing to our higher angular resolution and much larger map, we also observe differences in the morphology.


\section{Formation and excitation of CH$^+$ and OH via vibrationally excited H$_2$}
\label{sect:H2_comparison}

In this section, we compare the spatial distribution of the CH$^+$ J=3-2 line with those of H$_2$ v=1-0 S(1) and [CII] 158~$\muup$m, tracing the reactants of $C^+ + H_2 \to CH^+ + H$. We also compare the spatial distribution of the OH 84~$\muup$m emission to those of H$_2$ v=1-0 S(1) and [OI] 145~$\muup$m, the components of reaction $O + H_2 \to H + OH$.

\subsection{Spatial comparison of CH$^+$ with H$_2~(v=1)$ and C$^+$}

\begin{figure*}
	\centering
	\includegraphics[scale = 0.3]{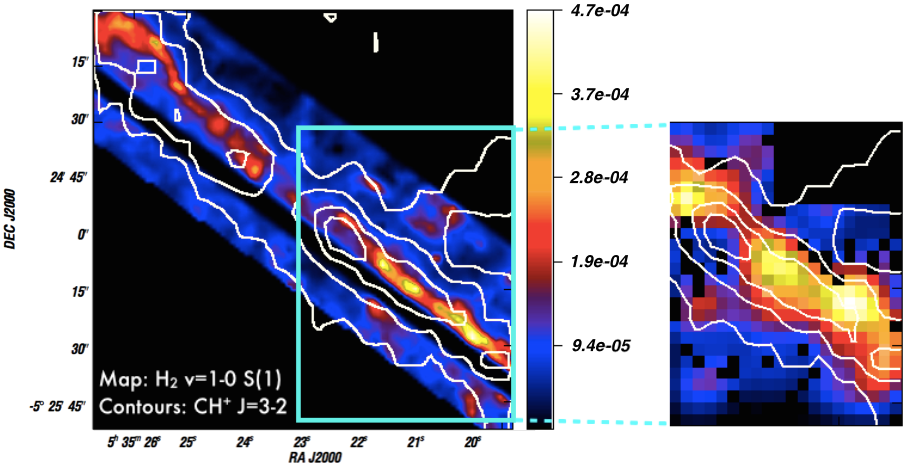} \hfill
	\includegraphics[scale = 0.24]{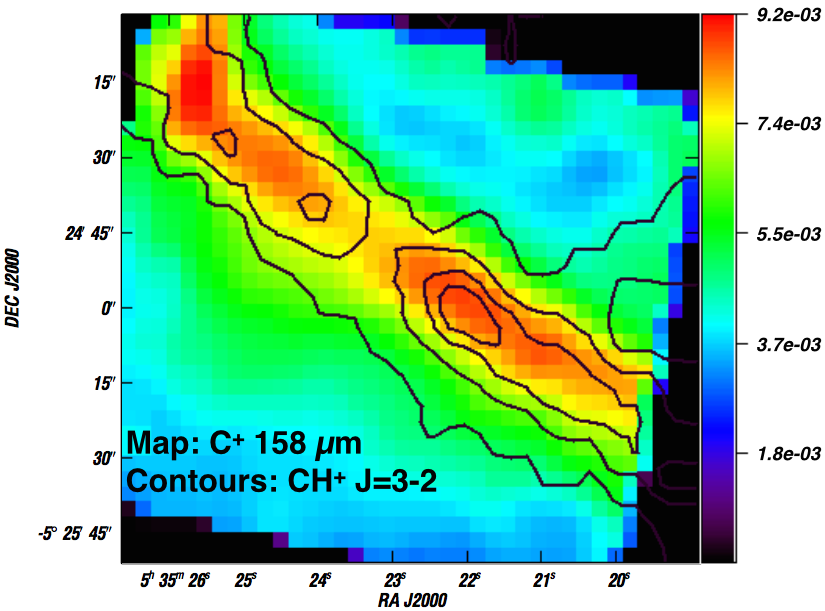}
	\includegraphics[scale = 0.3]{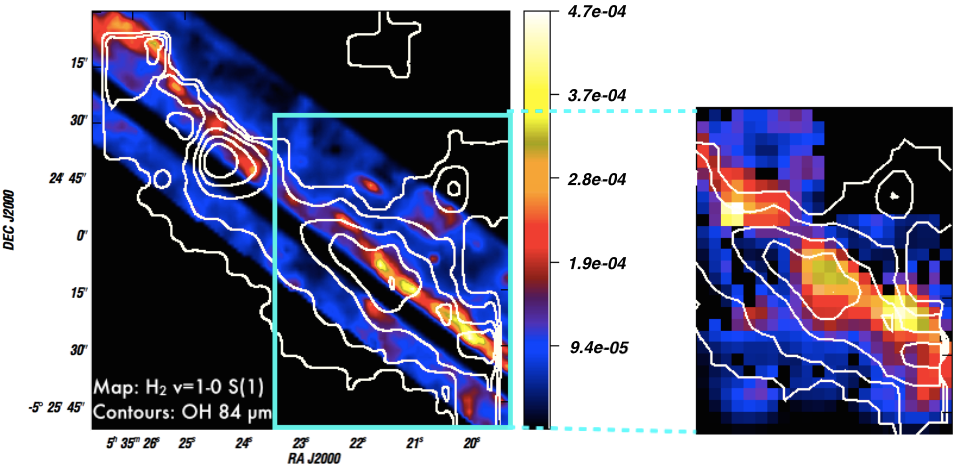} \hfill
	\includegraphics[scale = 0.24]{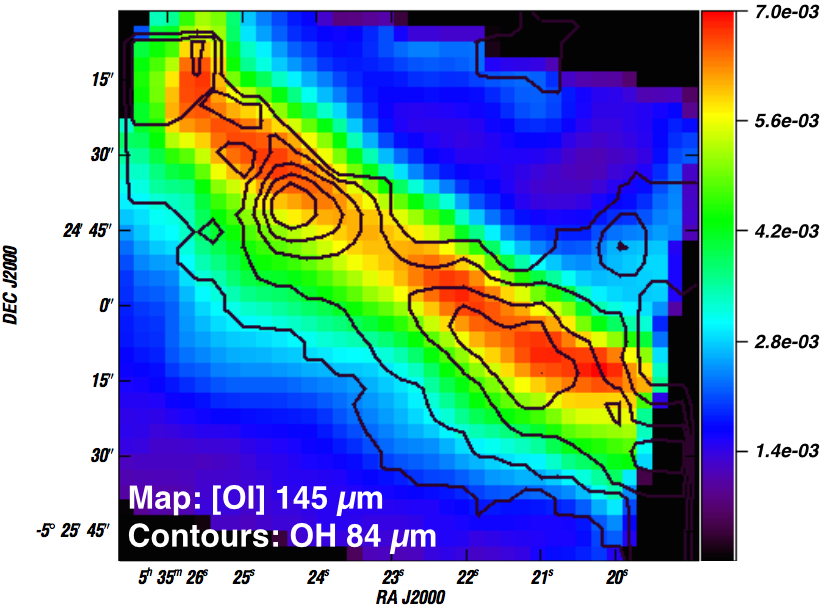}
	\caption [H2 and CHp]{Top row: CH$^+$ 120~$\muup$m compared with H$_2$ v=1-0 S(1)  and [CII] 158 $\muup$m. CH$^+$ 120~$\muup$m contours overlaid on the H$_2$ map (left) and [CII] map (right). Zoom: map of vibrationally excited H$_2$ convolved to the spatial resolution of CH$^+$ J=3-2 map including contours of CH$^+$ J=3-2.
	Bottom row: OH 84~$\muup$m compared with H$_2$ v=1-0 S(1) and [OI] 145~$\muup$m. OH 84~$\muup$m contours overlaid on the H$_2$ map (left) and [OI] map (right). 
All contours are with contour steps of 20~\%  of the peak emission, except for the zoomed picture where the contours are with contour steps of 10\%. The maps are in units of erg~s$^{-1}$~cm$^{-2}$~sr$^{-1}$.
The H$_2$ v=1-0 S(1) data taken from \citep[][big map]{Walmsley2000} and  \citep[][zoomed map]{VanderWerf1996}.
The [CII]~158 $\muup$m and [OI] 145~$\muup$m data were taken from \citep{Bernard-Salas2012}.}
    \label{pic:H2_comparison_maps}
\end{figure*}

Fig. \ref{pic:H2_comparison_maps} shows a map of H$_2$ v=1-0 S(1) at 2.12~$\muup$m and the CH$^+$ J=3-2 line. The H$_2$ v=1-0 S(1) was observed from the ground with a resolution of $\sim$1$\arcsec$ \citep{Walmsley2000}. The maps are not convolved, so that we do not blur the detailed structures. At the edge of the Bar, the contours of the CH$^+$ emission trace in detail the structures seen in H$_2$. In the same figure, we also include a smaller H$_2$ map \citep{VanderWerf1996} which is convolved to the same spatial scale. Again, we find a good correlation between the spatial distribution of the CH$^+$ J=3-2 emission and the vibrationally excited H$_2$ emission, except for clump 3 and in the lower bright structure seen in H$_2$. The correlation supports that CH$^+$ is formed by $C^+ + H_2 (v>0)$ and/or that the excitation of CH$^+$ is affected by formation pumping \citep{Agundez2010, Godard2013, Nagy2013, Zanchet2013}.

Fig. \ref{pic:H2_comparison_maps} also shows that CH$^+$ \mbox{J=3-2} correlates well with [CII] 158~$\muup$m \citep{Bernard-Salas2012}. Nevertheless, [CII] 158~$\muup$m is more extended in front and behind the Bar than CH$^+$ \mbox{J=3-2} (see Figs. \ref{pic:cuts} and \ref{pic:H2_comparison_maps}). As expected, C$^+$ exists in the outermost layers of the PDR, where most of the hydrogen is in atomic form and, thus, little CH$^+$ has formed. In the envelope surrounding the Bar, the physical conditions are still favorable for [CII] emission, but not for excited CH$^+$. The [CII] 158~$\muup$m line is easily excited because of its low excitation energy of 91~K. The background PDR contributes to 30\% of the [CII] emission \citep{Bernard-Salas2012}. The CH$^+$ J=3-2 emission is too faint to be detected from the background PDR.
 
Since the C$^+$ is more extended, it appears that the CH$^+$ emission is not limited by the C$^+$ abundance but by the abundance of vibrationally excited H$_2$. For PDRs with lower FUV radiation field, such as moderately excited PDRs (e.g., Horsehead or NGC~7023 East with radiation fields $\sim$100 times lower than in the Orion Bar) the CH$^+$ emission, and particularly the excited rotational transitions (J$>$2), is expected to decrease significantly. Towards the sample of PDRs observed by PACS and SPIRE as a part of the Evolution of Interstellar Dust key program \citep[][data available on the Herschel Idoc Database HESIOD webpage\footnote{\url{http://idoc-herschel.ias.u-psud.fr/sitools/client-user/}}]{Abergel2010}, [CII] 158~$\muup$m line is detected in all the PDRs while CH$^+$ J=1-0 line is only detected in PDRs associated to a large amount of vibrationally excited H$_2$. The line intensity upper limit for the non-detection of CH$^+$ (J=1-0) in the moderate excited PDRs, such as the Horsehead and NGC~7023 East, is 2$\times$10$^{-7}$~erg~s$^{-1}$~cm$^{-2}$~sr$^{-1}$ with SPIRE \citep[e.g., ][]{Kohler2014}. In the Horshead PDRs, the CH$^+$ (J=1-0) has not even been detected with HIFI within the rms limit of 4~mK. Low FUV radiation field and thermal pressure are clearly not favorable for the CH$^+$ line emission.

In conclusion, from our comparison of CH$^+$ with vibrationally excited H$_2$ and high-J CO, CH$^+$ J=3-2 emission is due to both formation pumping and collisions. This agrees with \citet{Godard2013} and \citet{Zanchet2013}, who found that formation pumping dominates the excitation of the $J \ge 2$ levels in the physical conditions of the Orion Bar.

\subsection{Spatial comparison of OH with H$_2~(v>0)$ and O$^0$}

In general, the OH 84~$\muup$m line map follows the H$_2$ emission (see Fig. \ref{pic:H2_comparison_maps}), but the OH emission is more extended and less correlated with H$_2 \, (v=1)$ than CH$^+$. An energy activation barrier exists in the $O + H_2$ reaction even for H$_2 \, (v=3)$ \citep{Sultanov2005, Agundez2010}, thus, the OH formation is expected to be less sensitive to the presence of H$_2$ ($v=$1).

In Fig. \ref{pic:H2_comparison_maps}, we also compare OH 84~$\muup$m with [OI] 145~$\muup$m line, the other component of the proposed formation route. Similarly to [CII], the [OI] emission is more extended and shifted by 5$\arcsec$ to the edge of the PDR compared to the OH emission, although the general features are similar. The differences can be explained by [OI] being present in the atomic PDR layers with low molecular gas fractions where H$_2$ is photodissociated and cannot participate in forming OH. The background PDR contributes to 15\% of the [OI] emission \citep{Bernard-Salas2012}. Like CH$^+$ \mbox{J=3-2}, we do not detect OH 84~$\muup$m line from the background PDR.

In conclusion, the spatial comparison of OH with [OI] and H$_2$ 1-0 S(1) indicates that the reaction $O^0 + H_2 \, (v=0)$ is the likely OH formation route at the physical conditions of the Orion Bar as predicted by \citet{Agundez2010}. Thus, OH excitation is more sensitive to the temperature and density variations than to the presence of H$_2$ (v$>$0).

\subsection{OH and H$_2$O distribution comparison}

\begin{figure*}
	\centering
	\includegraphics[scale = 0.25]{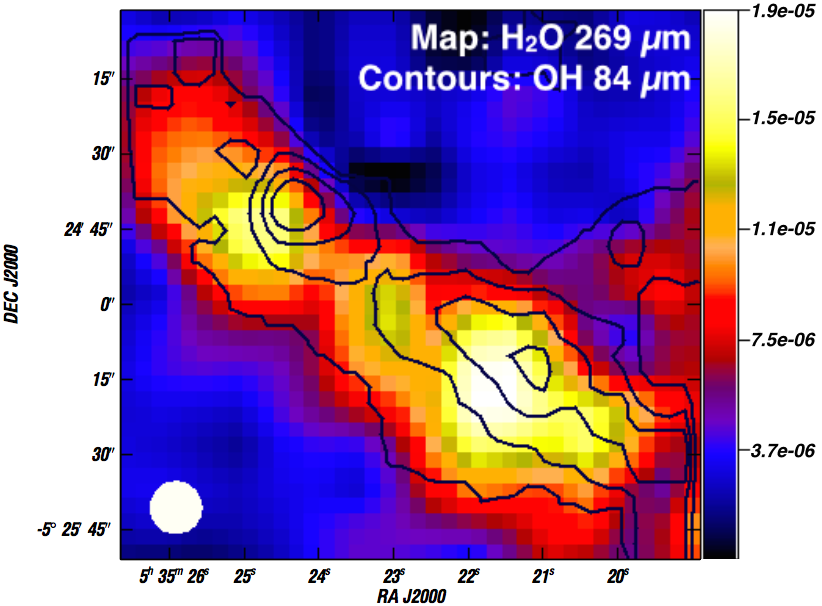}
	\qquad
	\qquad
	\includegraphics[scale = 0.25]{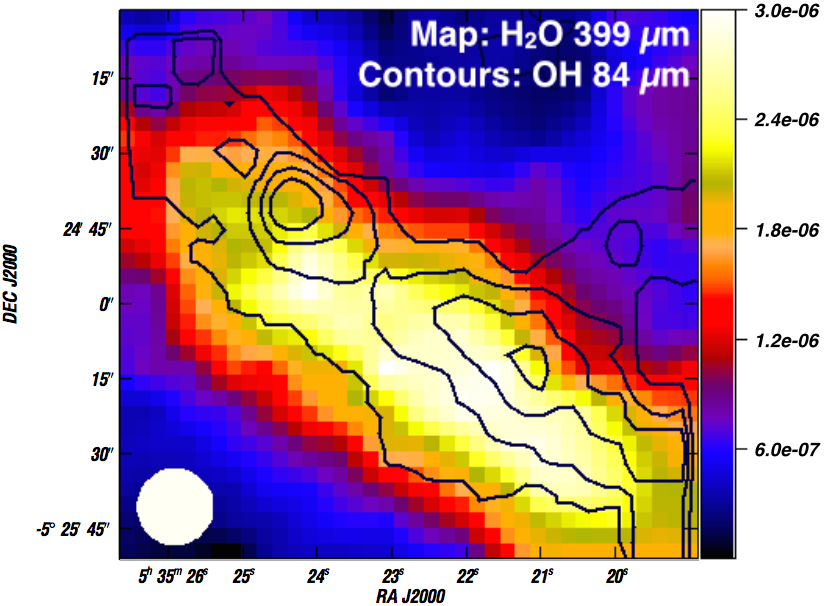}
	\caption[water]{Maps of H$_2$O 1$_{11}$-0$_{00}$ at 269 $\muup$m (left) and H$_2$O 2$_{11}$-2$_{02}$ 399 $\muup$m (right) with contours of OH 84.6 $\muup$m with contour steps of 20~\%  of the peak emission. The maps are in units of erg~s$^{-1}$~cm$^{-2}$~sr$^{-1}$.} 
    \label{pic:water_comparison}
\end{figure*}

In the warm gas unshielded against FUV radiation, there is an alternative formation route for OH via the photodissociation of H$_2$O \citep{Hollenbach2009, Hollenbach2012}. Photodissociation is the main destruction route of H$_2$O and OH.
The photodissociation is predicted to limit the H$_2$O abundance in the warm zone and the peak of H$_2$O emission is expected to be deeper in the PDR \citep{Hollenbach2009}. Here, we compare the OH~84~$\muup$m line with two water lines at 269 $\muup$m (\mbox{1$_{11}$-0$_{00}$}, E$_u$/k=53~K) and 399 $\muup$m (\mbox{2$_{11}$-2$_{02}$}, E$_u$/k=137~K) in Fig. \ref{pic:water_comparison}.

Both H$_2$O lines, more easily excited than OH~84~$\muup$m, peak behind the OH by $\sim$15$\arcsec$ and are more extended (see the cut in Fig. \ref{pic:cuts}). This could be explained by the lower gas temperature in the regions where the bulk of the H$_2$O arises compared to OH. The photodissociation is likely to limit the water abundance in the warm zone. Thus, the bulk of the OH and water emission arise from different depths. It also implies that this formation route is less important than the $O^0 + H_2 \to H + OH$ reaction.


\section{CH$^+$ kinematics}
\label{sect:HIFI} 

\begin{figure}
	\centering
	\includegraphics[scale = 0.2]{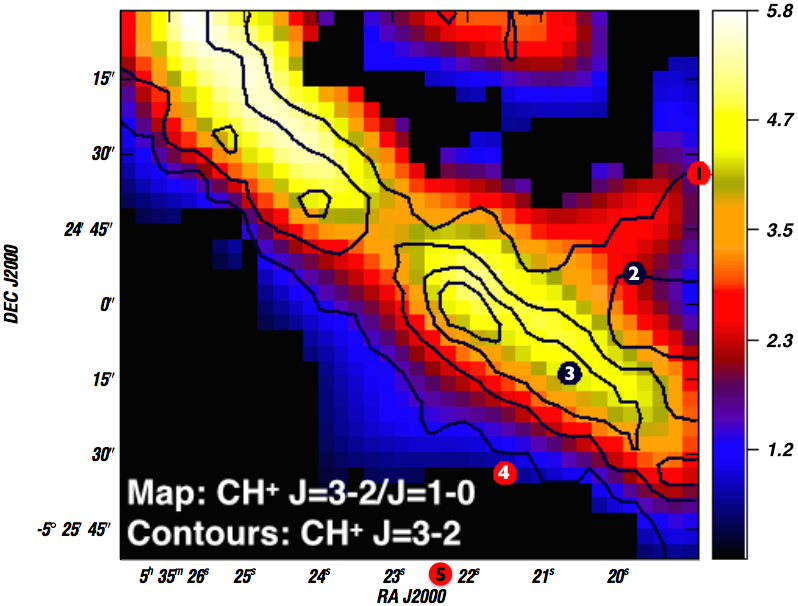}
	\includegraphics[scale = 0.3]{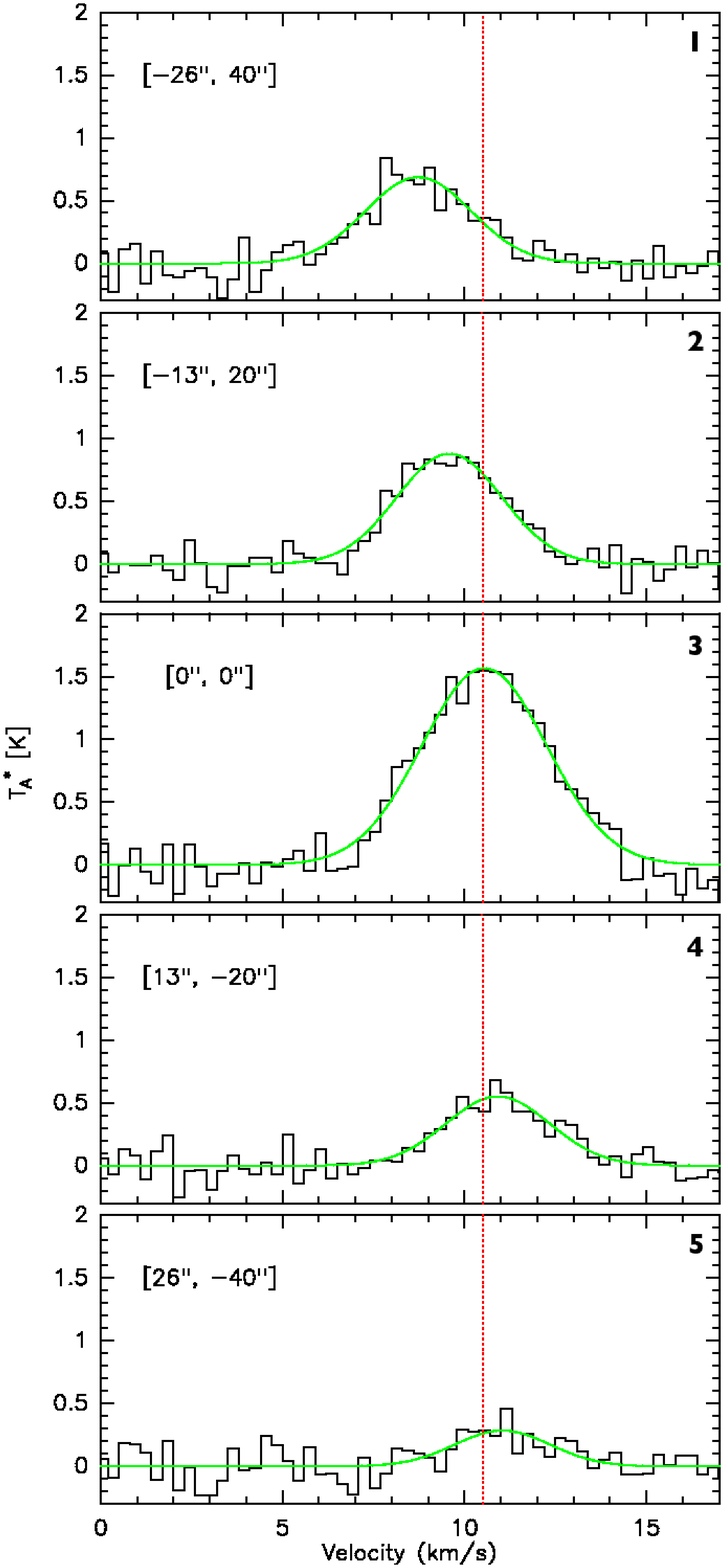}
	\caption [CHp excitation]{Upper panel: The ratio of CH$^+$ J=3-2 and CH$^+$ J=1-0 integrated intensity with contours of CH$^+$ J=3-2 in steps of 20~\% of the peak emission to indicate the position of the Bar. Lower panels: HIFI spectra of CH$^+$ J=2-1 at the positions across the Bar indicated in the map with corresponding numbering.}
    \label{pic:HIFI}
\end{figure}

To investigate gas kinematics and CH$^+$ line width variations, we examine the CH$^+$ J=2-1 (180~$\muup$m) line. We have velocity-resolved line profiles of a strip across the Southwestern part of the Bar observed with HIFI (see Fig. \ref{pic:config} for configuration). A selection of the spectra obtained along the cut are shown in Fig. \ref{pic:HIFI}.
The selected positions are $\sim$23$\arcsec$ apart from each other. 
The spectra have been obtained averaging all the scans that fall within 4$\arcsec$ of the selected position. 

The line centroid velocity varies from 9~km~s$^{-1}$ in front of the Bar to 11~km~s$^{-1}$ in the Bar and behind the Bar (see Table \ref{table:HIFI}). These centroid velocities are similar to what \citet{Peng2012} found for CO transitions from J=6-5 to J=8-7. We find a broad line width ranging from $\Delta V=3.4\pm0.3-3.1\pm0.7$ (in front and behind the Bar) to 4$\pm0.2$~km~s$^{-1}$ (in the Bar). This is consistent with the observations of \citet{Nagy2013} on the {\it CO$^+$ peak} who found line widths of $\Delta V=5.46\pm$0.04~km~s$^{-1}$ and $~4.57\pm$0.11~km~s$^{-1}$ for CH$^+$ J=1-0 and J=2-1, respectively. The CH$^+$ J=2-1 line appears narrower that the J=1-0 line which could be due to opacity broadening and to the larger beam size at the lower frequency. CH$^+$ has a broader line width than any other molecular tracer of dense gas in the Orion Bar (typically showing line widths of 2-3~km~s$^{-1}$, Joblin et al, in prep.). The CH$^+$ \mbox{J=2-1} line appears to be broad all over the region, even though the ratio of the CH$^+$ J=3-2 to J=1-0 shown in Fig. \ref{pic:HIFI} points to different physical conditions.

The broad width of CH$^+$ line could originate in a low density interclump component (n$\sim$10$^4$~cm$^{-3}$), where low density tracers such as [CII] have been observed to have similar line widths of $\sim$4~km~s$^{-1}$ \citep{Nagy2013}.
However, our maps show that the excited CH$^+$ and high-J CO both originate both in the dense and warm region of the Bar, while their line widths are different (3~km~s$^{-1}$ for the high-J CO, Joblin et al., in prep). Alternatively, based on reactivity of the molecule, \citet{Nagy2013} suggest that the broadening could be explained by formation pumping. If this is the case, there is no time for the CH$^+$ to thermalize, and the line width will remain unchanged. In conclusion, our results agree with \citet{Nagy2013} that the broad line width of CH$^+$ is most likely a signature of formation pumping.


\section{Proplyd in the Orion Bar}
\label{sect:proplyd}

\begin{figure}
	\centering
	\includegraphics[scale = 0.35]{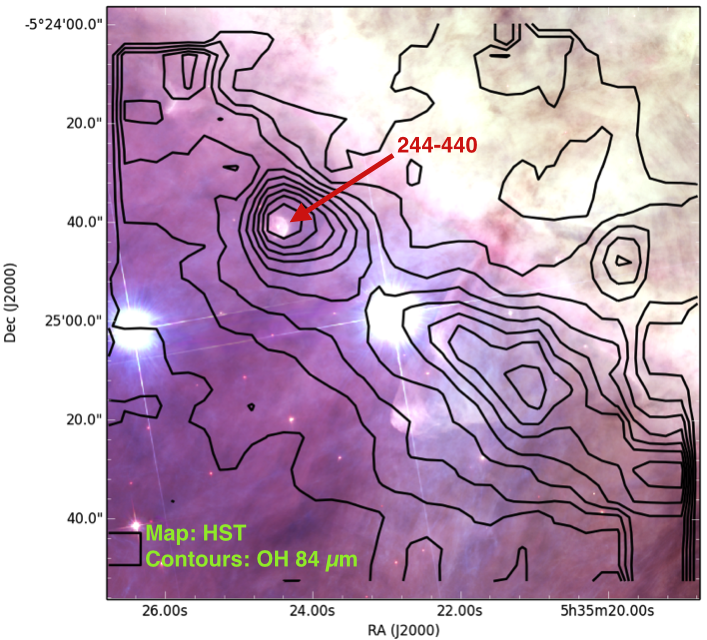}
	\caption [proplyd]{Contours of OH 84.6 $\muup$m at every 10 \% of the emission peak overlaid on an optical image from the HST \citep{Robberto2013}.}
    \label{pic:proplyd}
\end{figure}

The OH 84~$\muup$m map reveals a clear emission peak not detected in the other lines. In Fig. \ref{pic:proplyd}, we compare the OH 84 $\muup$m emission to an optical image from the Hubble Space Telescope (HST).
The HST image reveals an optical source at the OH peak position. This source has been identified as the proplyd 244-440 \citep[e.g.,][]{Bally2000}. Proplyds, or externally illuminated dense protoplanetary disks, are a special class of low-mass young stellar objects modulated by a strong, external UV field. They are found embedded within or near a HII region and are identified by their typical cometary photoionized envelopes.

This is, to our knowledge, the first time that OH emission has been tentatively associated to a proplyd, although OH has been observed in a protoplanetary disk in the DIGIT survey \citep{Sturm2010}. In Fig. \ref{pic:proplyd}, the highest OH contour level at the proplyd position is 8$\arcsec$ (our resolution being 6$\arcsec$) which corresponds to 3000~AU at a distance of 415~pc. While disks in proplyds are small (100$-$500~AU), their photoevaporating envelopes can extend up to a couple of 1000~AU. The OH 84~$\muup$m line emission at this position could come from UV-radiation heating or shocks. HST optical image of the proplyd 244-440 reveals the presence of microjets. Alternatively, in proplyds, the surface of the disk is illuminated and is, thus, similar to a very dense and warm PDR, and is also expected to produce FIR OH emission. If the OH origin is UV heating, the proplyd detection would confirm that the OH does originate in hot irradiated structures.

Like OH, CH$^+$ J=3-2 has been detected towards protoplanetary flared disks such as HD 100546 and HD 97048  \citep{Thi2011, Fedele2013}. In the upper layers of the disks, which are directly exposed to the stellar radiation field, the reaction $C^+ + H_2 \, (v>0) \to CH^+ + H$ is expected to be dominant. We do not see any excess of CH$^+$ J=3-2 in the Orion Bar, nor any excess of high-J CO emission. As opposed to OH, CH$^+$ and high-J CO lines are fainter in disks \citep[e.g.,][]{Fedele2013, Meeus2013} and their emission is likely overshadowed by the strong nebular emission in the region. 


\section{Summary and conclusions}
\label{sect:conclusions}

We have presented \textit{Herschel} fully sampled maps of the FIR rotationally excited emission lines of CH$^+$ and OH towards the Orion Bar over a large area ($\sim$110$\arcsec$ $\times$ 110$\arcsec$). The CH$^+$ and OH lines delineate the edge of the Bar indicating that they are good tracers of the warm molecular zone and very sensitive to the physical conditions. Our main results are summarized as follows.

\begin{enumerate}
\item We confirm the correlation of OH and CH$^+$ with high-J CO emission (a tracer of dense gas). We find that the spatial thickness of the observed line emission layers results from the Bar being tilted towards the observer, an effect already inferred in other studies. OH, CH$^+$, and high-J CO originate in the inclined thin irradiated surface of the Bar with embedded dense structures. The emission peaks of these lines are associated with the surface of the largest clumps (density enhancements) seen with high density tracers, such as HCN or CS. 

\item Although excited CH$^+$ and OH display similar overall spatial distribution, there are also relevant differences in their morphology. The good correlation of CH$^+$ with vibrationally excited H$_2$ supports that CH$^+$ is formed via the reaction $C^+ \, + \, H_2 \, (v > 0)$. We provide observational evidence to the fact that the excitation of the CH$^+$ rotational levels is affected by formation pumping which dominates the excitation for J$\ge$2 levels, as suggested by previous models. The CH$^+$ emission is limited by the abundance of vibrationally excited H$_2$ and not by the C$^+$ abundance. We find the C$^+$ emission to be more extended than the CH$^+$ emission. 

\item OH is less well correlated with the vibrationally excited H$_2$, a fact pointed out by theoretical studies predicting the reaction $O^0 \, + \, H_2 \, (v=0) \to H + OH$ to dominate the OH formation. The other formation route for OH via the photodissociation of H$_2$O appears not to be as important since the bulk of the OH and water emission seem to arise from different cloud depths.

\item We find a broad line width of $\sim$3$-$4~km~s$^{-1}$ for CH$^+$ \mbox{J=2-1}, which is broader than other molecular line tracing dense gas in the Orion Bar. The line width most likely reflects the excess energy transferred to translational energy during the formation pumping of CH$^+$.

\item Interestingly, the peak of the OH 84 $\muup$m emission corresponds to the position of the proplyd 244-440. FIR OH emission could be a good diagnostic of hot gas in the externally illuminated protoplanetary disks.
\end{enumerate}


\begin{acknowledgements}
J. Bernard-Salas wishes to acknowledge the support of a Career Integration Grant within the 7th European Community Framework Program, FP7-PEOPLE-2013-CIG-630861-SYNISM.
J. R. Goicoechea thanks Spanish MINECO for funding support under grants CSD2009-00038 and AYA2012-32032. He also thanks the ERC for support under grant ERC-2013-Syg-610256-NANOCOSMOS. 
M. Gerin and B. Godard thank the French PCMI program for funding.
\end{acknowledgements}


\bibliographystyle{aa}
\bibliography{AA201629445}


\end{document}